\documentclass[preprint,12pt]{elsarticle}

\usepackage{amsmath}
\usepackage{amssymb}
\usepackage{graphics} 

\newcommand{\abs}[1]{\left|  #1  \right|}
\newcommand{\expv}[1]{\langle  #1 \rangle}
\newcommand{\rb}{\mathbf{r}}
\newcommand{\pb}{\mathbf{p}}
\newcommand{\rbb}{\mathbf{\overline{r}}}
\newcommand{\pbb}{\mathbf{\overline{p}}}

\newcommand{\laplace}{\Delta}
\newcommand{\pot}{V_\psi}
\newcommand{\dd}{\mathrm{d}}

\newcommand{\Fb}{\mathbf{F}}
\newcommand{\Dr}{\Delta\mathrm{r}}
\newcommand{\Dp}{\Delta\mathrm{p}}

\journal{Physics Letters A}

\begin{document}

\begin{frontmatter}

\title{Schr\"odinger-Newton equation with complex Newton constant and induced gravity}
\author[label1]{Lajos Di\'osi\corref{cor1}}
\cortext[cor1]{Corresponding author.}
\ead{diosi@rmki.kfki.hu}
\ead[url]{www.rmki.kfki.hu/~diosi}
\author[label2]{Tibor Norbert Papp}

\address[label1]{Research Institute for Particle and Nuclear Physics\\
H-1525 Budapest 114, POB 49, Hungary}
\address[label2]{Department of Physics of Complex Systems, E\"otv\"os University\\
H-1518 Budapest, POB 32, Hungary}
\ead{tibpap@ludens.elte.hu}

\begin{abstract}
In the reversible Schr\"odinger-Newton equation a complex Newton coupling $G\exp(-i\alpha)$ is proposed in place of $G$. 
The equation becomes irreversible and all initial one-body states are expected to converge to solitonic stationary states. 
This feature is verified numerically. For two-body solutions we point out that an effective Newtonian interaction is 
induced by the imaginary mean-fields as if they were real. The effective strength of such induced gravity depends on the 
local wave functions of the participating distant bodies. 
\end{abstract}


\end{frontmatter}

\section{Introduction}
The Schr\"odinger-Newton equation (SNE) was proposed in the context of
quantum foundations \cite{Dio84,Pen98}. Assuming a Newtonian self-interaction, 
it realizes the concept that gravity is responsible for 
the observed spatial localization of macroscopic objects. 
It leads to plausible scales of localization.  
The SNE has been studied in numerous works 
\cite{Moretal98,Beretal98,KumSon00,Tod01,Son02,GuzUre03,Ges04,SalCar07,GreWun06,Adl07,Dio07}.
Its basic feature is that the Newtonian mean-field self-interaction modifies the free Schr\"odinger
equation in such a way that it acquires localized ground states (solitons). These solutions
are then considered as natural `pointer' states for the macroscopic objects.
On the other hand, these stationary pointer states are expected to emerge through the evolution
of the wave function. Unfortunately, this irreversible mechanism can not be realized by the
reversible SNE. To make the solutions converge toward pointer states, one gives the Newton coupling $G$ 
a negative imaginary part. 
It turns the reversible SNE into an irreversible frictional SNE (frSNE). 
The specific frSNE with $G\rightarrow-iG$ has been an emergent structure in the author's decoherence model \cite{Dio87,Dio89},
and its relationship to the reversible SNE has been analyzed in Refs.~\cite{Dio07} and \cite{WezBri08}.
We recall the one-body frSNE in Sec.~\ref{One}, we solve it numerically and determine
the unique stationary wave packet in Sec.~\ref{Num}. We discuss the two-body frSNE in the special case when the bodies are
far form each other and find that the imaginary coupling induces gravitational
attraction, as if it were real coupling (Sec.~\ref{Two}). The strength of this induced gravity, however, turns out to depend
on the quantum state of the participating bodies. The general case of complex coupling 
$G\mathrm{e}^{-i\alpha}$ with $0<\alpha<\pi$ is briefly discussed in Sec.~\ref{Complex}. 

\section{One body equation}\label{One}
The SNE is a Schr\"odinger equation with a Newtonian mean-field potential where - in contrast to
textbook many-body equations - we retain the self-interaction terms. Therefore the mean-field potential  
is already present in the one-body SNE:
\begin{equation}\label{SNE}
\frac{\dd\psi}{\dd t} =\frac{i\hbar}{2M}\laplace\psi-\frac{i}{\hbar}\pot\psi~,
\end{equation}
where $\psi=\psi(\rb)$ is the c.o.m. wave function of the object of mass $M$, $G$ is the Newton constant, and
\begin{equation}\label{pot}
\pot=\pot(\rb)=-GM^2\int\frac{\abs{\psi(\rb')}^2} {\abs{\rb-\rb'}}\dd\rb'
\end{equation}
is the Newtonian mean-field potential of a point-like object. The stationary ground states are solitons of 
spread $(\Dr)_0\sim(\hbar^2/GM^3)$ \cite{Dio84,Pen98} but they do not attract other solutions.
We create a basin of attraction (convergence) if we give $G$ an imaginary part. Robust convergence
is achieved by the replacement $G\rightarrow-iG$:    
\begin{equation}\label{frSNE}
\frac{\dd\psi}{\dd t} =\frac{i\hbar}{2M}\laplace\psi-\frac{1}{\hbar}(\pot-\expv{\pot})\psi~.
\end{equation}
The imaginary mean-field corrupts the normalization of $\psi$, we restore it by the constant counter term:  
\begin{equation}\label{scalpot}
\expv{\pot}=-GM^2\int\frac{\abs{\psi(\rb)}^2\abs{\psi(\rb')}^2} {\abs{\rb-\rb'}}\dd\rb\dd\rb'~.
\end{equation}
The frSNE preserves both the momentum and the position 
expectation values $\expv{\pb}\equiv\pbb$ and $\expv{\rb}\equiv\rbb$. If, for simplicity, we start a wave function with 
$\pbb=0$ and $\rbb=0$ then the solution will converge to the rotational invariant stationary state
\begin{equation}\label{stac_sol_spher}
\psi_0(r)\mathrm{e}^{-(i/\hbar)E_0 t}~,
\end{equation} 
where
\begin{equation}\label{E_stac}
E_0 =\frac{\hbar^2}{2M}\int\limits_0^\infty\abs{\psi_0'(r)}^2 4\pi r^2\dd r~.
\end{equation}
Surprisingly, the `energy' $E_0$ of the stationary state (\ref{stac_sol_spher}) contains the kinetic energy only, 
without the contribution of the Newton self-interaction. This is a consequence of the choice $G\rightarrow-iG$ which makes the mean-field pure
imaginary, there is no real dynamical potential left in the frSNE. 

Little is known about the details of the stationary solution (\ref{stac_sol_spher}). Dimensional analysis of the standard 
spread yields the same order of magnitude $(\Dr)_0\sim(\hbar^2/GM^3)$ as in case of the reversible SNE \cite{Dio84}.
Analytic results exist for large extended
spherical objects \cite{Dio89,Dio07}; for point-like bodies we have to use numeric simulations. Once we know the stationary
solution $\psi_0$ in the c.o.m. frame then, thanks to the Galilean invariance of the frSNE, we know all stationary solutions: 
\begin{equation}\label{stac_sol}
\psi_0(\abs{\rb-\rbb_t})\mathrm{e}^{(i/\hbar)(\pbb\rb-Et)}~,
\end{equation}
where $E=E_0+(\pbb^2/2M)$ and $\dd\rbb_t/\dd t=\pbb/M$ while $\pbb=\mathrm{const}$. If the frSNE is perturbed by a weak smoothly varying
field, real or imaginary, we can still retain the above form if we make the c.o.m. momentum $\pbb$ time-dependent. This will be the case later
when we apply the frSNE to two bodies far from each other. The following position-momentum correlation matrix will play a definitive
role in their effective interaction: 
\begin{equation}\label{R}
\mathsf{R} \equiv \frac{1}{\hbar}\mathrm{Re}\left(\expv{\pb\circ\rb}-\expv{\pb}\circ\expv{\rb}\right)
=\frac{-i}{2}\int\psi^\ast(\rb)(\nabla\circ\rb+\rb\circ\nabla)\psi(\rb)\dd\rb-\frac{1}{\hbar}\pbb\circ\rbb~.
\end{equation}
Due to the rotational invariance of $\psi_0$, the stationary matrix $\mathsf{R}_0$ becomes proportional to the unit matrix $\mathsf{I}$,
i.e.: $\mathsf{R}_0=R_0\mathsf{I}$. The correlation scalar $R_0$ will be determined numerically. 

\section{Numerical solution}\label{Num}
\begin{figure}
\centering
\resizebox{80mm}{!}{\rotatebox{-90}{\includegraphics{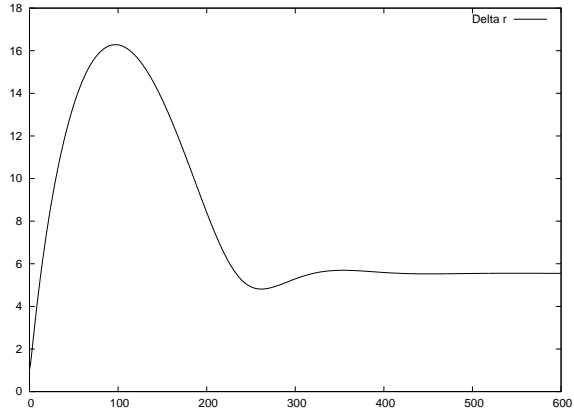}}} 
\caption{Relaxation of the standard spread $\Dr$ of the wave function $\psi(r,t)$
toward the stationary value $(\Dr)_0=5.5501$ in function of time. A real Gaussian of spread $1$ was the initial wave function ($\hbar=G=M=1$).}
\label{fig:dev}
\end{figure}

\begin{figure}
\centering
\resizebox{80mm}{!}{\rotatebox{-90}{\includegraphics{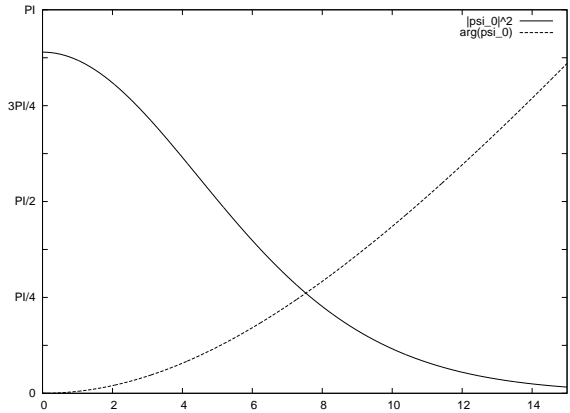}}} 
\caption{Spatial density $\vert\psi_0(r)\vert^2$ (full) and phase $\chi_0(r)$ (dashed), resp., of the stationary wave function $\psi_0(r)$ in function
of the radial coordinate $r$. Standard position spread is $(\Dr)_0=5.5501$, momentum spread is $(\Dp)_0=0.2668$, 
position-momentum correlation is $R_0=0.6753$  ($\hbar=G=M=1$).}
\label{fig:abs2}
\end{figure}

We restricted our numeric simulations for rotational invariant states $\psi(\rb) = \psi(r)$. 
The mean-field (\ref{pot}) takes the following form:
\begin{equation}\label{pot_spher}
\pot(r)=-4\pi GM^2\int\limits_0^\infty  
\frac{\abs{\psi(r')}^2}{\max(r,r')}r'^2 \dd r'.
\end{equation}
The frSNE (\ref{frSNE}) reduces to:  
\begin{equation}\label{frsne_spher}
\frac{\dd\psi}{\dd t} =\frac{i\hbar}{2M}\left(\frac{2}{r}\psi' + \psi''\right) 
-\frac{1}{\hbar}(\pot-\expv{\pot})\psi~.
\end{equation}
We have simulated the solution of this equation with various initial wave functions (Gaussian, smoothened rectangle,
superposition of two Gaussians). They all converged to the unique localized stationary solution (\ref{stac_sol_spher}).
We illustrate the convergence process by monitoring the time-dependence of the standard spread $\Dr$ of the wave function:  
\begin{equation}
(\Dr)^2 = \frac{4\pi}{3}\int\limits_0^\infty |\psi(r)|^2r^4\dd r~. 
\end{equation}
A competition between the spreading kinetic term and the contractive potential term is witnessed by  
transient oscillations until the stationary value $(\Dr)_0=5.5501(\hbar^2/GM^3)$ is reached (Fig.~\ref{fig:dev}).
The Fig.~\ref{fig:abs2} shows the stationary distribution $\vert\psi_0(r)\vert^2$.
It is bell-shaped but it is not Gaussian. Also the complex phase of the numeric solution $\psi_0(r)$ is shown in the same figure.
We numerically observed the constant shift of the phase at speed $-E_0/\hbar$, with $E_0=0.0356(G^2M^5/\hbar^2)$. This value has independently
been confirmed by numeric integration of the expression (\ref{E_stac}) using the
numerically calculated $\psi_0$.
We also determine the stationary position-momentum correlation matrix (\ref{R}) in the numerically obtained stationary state $\psi_0$. 
The correlation scalar $R_0$ can be written as: 
\begin{equation}\label{R_spher}
R_0 = \frac{4\pi}{3\hbar}
\int\limits_0^\infty r^3 \vert\psi_0(r)\vert^2 \chi_0'(r) \dd r~,
\end{equation}
where $\chi_0=\arg(\psi_0)$. Numeric integration obtained $R_0=0.6753$. 

\section{Two body equation}\label{Two}
For two identical objects the wave function reads $\Psi(\rb_1,\rb_2)$ and
the frSNE has the following structure \cite{Dio89,Dio07}: 
\begin{eqnarray}\label{frSNE2}
\frac{\dd\Psi}{\dd t}&=&\frac{i\hbar}{2M}\laplace_1\Psi-\frac{1}{\hbar}(\pot^{(11)}+\pot^{(12)}-\expv{\pot^{(11)}+\pot^{(12)}})\Psi\nonumber\\
       &\!&\!\!\!\!\!\!+\frac{i\hbar}{2M}\laplace_2\Psi-\frac{1}{\hbar}(\pot^{(22)}+\pot^{(21)}-\expv{\pot^{(22)}+\pot^{(21)}})\Psi~,
\end{eqnarray}
where the Laplacians $\laplace_1,\laplace_2$ refer to $\rb_1,\rb_2$, respectively. The mean-field self-interactions read:
\begin{eqnarray}\label{pot_carte_1122}
\pot^{(11)}(\rb_1)&=&-GM^2\int\frac{\abs{\Psi(\rb_1',\rb_2')}^2}{\abs{\rb_1-\rb_1'}}\dd\rb_1' \dd\rb_2'~,\\
\pot^{(22)}(\rb_2)&=&-GM^2\int\frac{\abs{\Psi(\rb_1',\rb_2')}^2}{\abs{\rb_2-\rb_2'}}\dd\rb_1' \dd\rb_2'~,
\end{eqnarray}
the mean-field cross-interactions read:
\begin{eqnarray}\label{pot_carte_1221}
\pot^{(12)}(\rb_1)&=&-GM^2\int\frac{\abs{\Psi(\rb_1',\rb_2')}^2}{\abs{\rb_1-\rb_2'}}\dd\rb_1' \dd\rb_2'~,\\
\pot^{(21)}(\rb_2)&=&-GM^2\int\frac{\abs{\Psi(\rb_1',\rb_2')}^2}{\abs{\rb_2-\rb_1'}}\dd\rb_1' \dd\rb_2'~.
\end{eqnarray}
 
We do not intend to discuss the generic two-body solutions. We are interested in the special case when the two bodies are
far from each other at locations $\rbb_1=\expv{\rb_1}$ and  $\rbb_2=\expv{\rb_2}$. If
$\abs{\rbb_1-\rbb_2}$ is much larger than the one-body stationary extension $(\Dr)_0$ then we expect the following
scenario. Considering first the limit $\abs{\rbb_1-\rbb_2}\rightarrow\infty$, 
the cross-interactions $\pot^{(12)}$ and $\pot^{(21)}$ can be omitted,
the two-body frSNE (\ref{frSNE2}) splits into two separate one-body equations (\ref{frSNE}): 
both bodies reach their localized stationary states in their respective c.o.m. frames.
Accordingly, the two-body solution is simply the product 
\begin{equation}\label{psi_stac2}
\Psi(\rb_1,\rb_2,t)=\psi_0^{(1)}(\rb_1,t)\psi_0^{(2)}(\rb_2,t)
\end{equation}
of the one-body solutions (\ref{stac_sol}):
\begin{equation}      
\psi_0^{(1)}(\rb_1,t)=\psi_0(\abs{\rb_1-\rbb_{1t}})\mathrm{e}^{(i/\hbar)(\pbb_1\rb_1-E_1 t)}~,
\end{equation}
where $E_1=E_0+(\pbb_1^2/2M)$ and $\dd\rbb_{1t}/\dd t=\pbb_1/M$, and a similar definition holds for $\psi_0^{(2)}$, too. 
As we said, the
effect of the cross-interaction terms $\pot^{(12)},\pot^{(21)}$ is perturbative and leads to the slight acceleration of the
c.o.m. momenta $\pbb_1,\pbb_2$. Let us calculate the acceleration of $\pbb_1$. First, we expand the cross-interaction term
$\pot^{(12)}$:
\begin{equation}\label{potlin}
\pot^{(12)}(\rb_1)=\pot^{(12)}(\rbb_1)-\Fb(\rb_1-\rbb_1)~,
\end{equation}
where:
\begin{equation}\label{F12}
\Fb=GM^2\frac{\rbb_2-\rbb_1}{{\abs{\rbb_1-\rbb_2}}^3}~.
\end{equation}
If we substitute the ansatz (\ref{psi_stac2}) into the two-body frSNE (\ref{frSNE2}), the following separate structure can be
obtained for $\psi_0^{(1)}$:
\begin{equation}\label{frSNE1}
\frac{\dd\psi_0^{(1)}}{\dd t}=\frac{1}{\hbar}\Fb(\rb_1-\rbb_1)\psi_0^{(1)}~.
\end{equation}
This is an effective frSNE, valid for $(\Dr)_0\ll\abs{\rbb_1-\rbb_2}$, 
to calculate the c.o.m. acceleration $\dd\pbb_1/\dd t$ of the stationary wave packet of the
body. The vector $\Fb$ is the Newton force at location $\rbb_1=\expv{\rb_1}$ caused by the other body at the remote location  
$\rbb_2=\expv{\rb_2}$, and $-\Fb$ is the force on the second body caused by the first one. 
A similar equation could be derived for $\dd\pbb_2/\dd t$, with the opposite force $-\Fb$. 
However, the vector $\Fb$ \emph{does not play the role of a real force} in the Eq.~(\ref{frSNE1}). 
If it did, it should have come with the additional imaginary factor $i$. 
Nevertheless, we are going to prove that even these imaginary Newton forces can mimic the 
true Newton forces. Indeed, they do accelerate the bodies in a Newtonian way apart from a numeric factor $R_0$. 
The acceleration $\dd\pbb_1/\dd t$ has the standard form:
\begin{equation}
\frac{\dd\pbb_1}{\dd t}=-i\hbar\int[\psi_0^{(1)}(\rb_1)]^\ast\nabla_1\frac{\dd\psi_0^{(1)}(\rb_1)}{\dd t}\dd\rb_1
                               -i\hbar\int\left[\frac{\dd\psi_0^{(1)}(\rb_1)}{\dd t}\right]^\ast\nabla_1\psi_0^{(1)}(\rb_1)\dd\rb_1~.
\end{equation} 
Let us substitute the effective equation (\ref{frSNE1}) and recall the matrix $\mathsf{R}_0$ for the
position-momentum correlation (\ref{R}) in the stationary state $\psi_0$. 
After trivial steps we get $\dd\pbb_1/\dd t=2\mathsf{R}_0\Fb$ which, due to the rotational symmetry, amounts to the ultimate form:
\begin{equation}
\frac{\dd\pbb_1}{\dd t} = 2R_0\Fb\approx1.3506\Fb~,  
\end{equation}
and, of course, similar steps would give the opposite acceleration $\dd\pbb_2/\dd t=-2R_0\Fb\approx-1.3506\Fb$ 
for the other body. The value of $R_0$ was obtained numerically in Sec.~\ref{Num}.

\section{Complex coupling}\label{Complex}
\begin{figure}
\centering
\resizebox{80mm}{!}{\rotatebox{-90}{\includegraphics{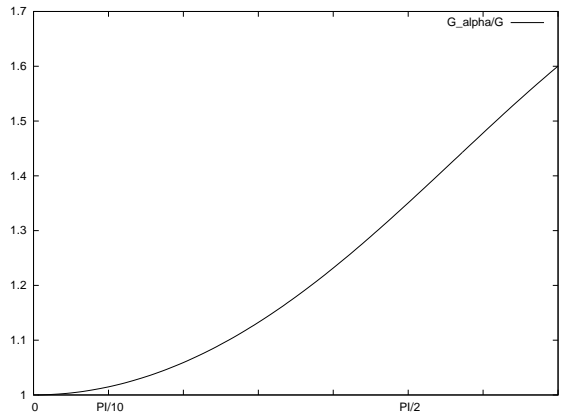}}} 
\caption{The ratio of the effective coupling $G_\alpha$ to Newton's $G$ in function of the phase $\alpha$ of the complex Newton 
coupling $G\mathrm{e}^{-i\alpha}$.}
\label{fig:Geff}
\end{figure}
We outline some features of the general frSNE - first advocated in the present
Letter - where the coupling $G$ in the SNE is replaced by $G\mathrm{e}^{-i\alpha}$. 
Obviously, $\alpha=0$ means the reversible SNE (\ref{SNE}), $\alpha=\pi/2$ is the case
$G\rightarrow-iG$ known earlier \cite{Dio89,WezBri08} and studied through Secs.~\ref{One}-\ref{Two} 
while $\alpha=\pi$ would mean the reversible SNE with repulsive Newton gravity
which we exclude from considerations. Inside the region $0<\alpha<\pi$ we have generalized frSNEs with features
resembling the investigated special case $\alpha=\pi/2$. All these frSNEs have localized one-body asymptotic states of unique
shape with parameters of the same order of magnitude. Convergence was numerically observed already at small imaginary 
couplings. (We conjecture that the ground state solution of the SNE is exactly indetical
with the stationary state $\psi_0$ of the frSNE with coupling $G-i\epsilon$, where $\epsilon\rightarrow+0$.) 
For $\alpha=\pi/2$ one has a fairly robust convergence which becomes finally lost 
towards the edges $\alpha=0,\pi$. For a general frSNE, the strength of Newton attraction has two contributions: the true
dynamics which is proportional to $\mathrm{Re}G$ and the induced acceleration which is proportional to $2R_0\mathrm{Im}G$.
The resulting gravity is characterized by an effective coupling: 
\begin{equation} 
G_\alpha=(\cos\alpha + 2R_0 \sin\alpha)G~.
\end{equation}
On Fig.~\ref{fig:Geff}, we have plotted the factor $\cos\alpha + 2R_0 \sin\alpha$ in function of the phase $\alpha$.
We see that the effective Newton forces are bigger than the standard ones by a small numeric factor about, apparently, less than $2$.
We could renormalize the bare coupling $G$ by the inverse of this factor in order to get the right effective
coupling. This discussion is beyond our scope now, yet we mention a very similar issue with an alternative - and perhaps related -
concept of emergent (induced) gravity \cite{Dio09}.  

\section{Summary}\label{Sum}
We have studied the simplest frictional Schr\"odinger-Newton equation characterized by the pure imaginary coupling $-iG$. 
We solved the one-body frSNE numerically and found robust convergence to a unique localized wave packet in the c.o.m. frame. 
In case of the two-body frSNE we discussed a heuristic solution in the special case when the two bodies are far from each other.
They quickly form stationary wave packets in their respective c.o.m. frames. An effective attraction emerges between them
which is $2R_0$ times the Newtonian attraction where $R_0$ is the position-momentum correlation of the one-body stationary
wave packet. Therefore our `induced gravity' depends on the details of the wave function of the participating bodies.     
Before they reach their stationary states towards the limit $2R\rightarrow2R_0\approx1.3506$, 
the emergent gravity may be very different from Newton's. As to the stationary regime itself, an immediate question arises: 
can we tune the factor $R_0$ to $1/2$? Yes, we can. The frSNE of extended spherical bodies yields just $R_0=1/2$ \cite{Dio89,Dio09}. 
Therefore it may well be that the concept of the frSNE with pure imaginary coupling $-iG$ contains a bit of real physics:
for large extended objects the frSNE would show stable convergence to one-body localized states and induces correct Newton forces 
between the distant objects. Yet, we would not like to over-interprete this theory since any SNE in itself faces serious 
interpretational problems, like any other non-linear deterministic Schr\"odinger equation \cite{Gis90}. Finally, we discussed
a novel class of frSNE with general complex coupling. Our simulations show that for point-like objects the effective 
(real + induced) gravity is bigger than Newton's by a small factor. 

In both our conceptional and numerical analysis we followed restricted aims. Without sinking into the interpretational
context (wave function collapse, macroscopic quantum mechanics, quantum gravity, etc.) we wanted to point out the hitherto 
unknown mechanism of gravity induced by the imaginary mean-field.
Our numeric simulations served this aim basically, we have not targeted a systematic numeric study of the frSNE.    
We add, nonetheless, the irreversible frSNE is an easy subject of simulation, compared to the irreversible SNE,
its numeric study might be an attractive further task in itself while, more importantly, it would serve our conceptional
understanding a possible role of the frSNE in foundations.
     
\section*{Acknowledgment}
This work was supported by the Hungarian OTKA Grants No. 49384, 75129.

\end{document}